\documentclass[fleqn,usenatbib]{mnras}

\usepackage{newtxtext,newtxmath}

\usepackage[T1]{fontenc}

\DeclareRobustCommand{\VAN}[3]{#2}
\let\VANthebibliography\thebibliography
\def\thebibliography{\DeclareRobustCommand{\VAN}[3]{##3}\VANthebibliography}

\usepackage{graphicx}	
\usepackage{amsmath}	
\usepackage{lineno}
\usepackage{bm}
\usepackage{enumitem}
\usepackage{makecell}
\hypersetup{
	colorlinks = true,
	linkcolor = blue,
	anchorcolor = blue,
	citecolor = blue,
	urlcolor = blue
}
\usepackage[caption=false]{subfig}
\usepackage{soul, color, xcolor}
\usepackage{multirow}
\usepackage{booktabs}
\usepackage{threeparttable}

\title[Discovery of high-frequency QPO in short GRBs]{Discovery of high-frequency quasi-periodic oscillation in short-duration gamma-ray bursts}

\author[Yang et al.]
{Xing Yang$^{1}$, Hou-Jun L\"{u}$^{1}$\thanks{E-mail: lhj@gxu.edu.cn}, Jared Rice$^{2}$, and En-Wei Liang$^{1}$  \\
 $^1$Guangxi Key Laboratory for Relativistic Astrophysics, School of Physical Science and Technology, Guangxi University, Nanning 530004, China\\
 $^{2}$Department of Mathematics and Physical Science, Southwestern Adventist University, Keene, TX 76059\\}

\begin{document}
\label{firstpage}
\pagerange{\pageref{firstpage}--\pageref{lastpage}}
\maketitle

\begin{abstract}
Rapidly rotating newborn magnetars, which originate from binary neutron star (NS) mergers and serve as the central engines of short gamma-ray bursts (GRBs), may leave some imprints on their prompt gamma-ray light curves even though they are far from their radiating fireballs. A high-frequency quasi-periodic oscillation (QPO) would be a unique feature for the magnetar central engine, especially a hypermassive magnetar. By conducting a systematic analysis of the prompt gamma-ray light curves from 605 short GRBs observed by {\em Fermi}/Gamma-ray Burst Monitor, we have identified such QPO signals in three GRBs (e.g. GRB 120323A, GRB 181222B, and GRB 190606A). The QPOs that peaked at $1258^{+6}_{-6}$ Hz for GRB 120323A, $623^{+4}_{-4}$ Hz for GRB 181222B, and $1410^{+4}_{-5}$ Hz for GRB 190606A are all with a confidence level above 5.2 $\sigma$. The high-frequency QPO signals of those three short GRBs may be caused by a hypermassive magnetar acting as the central engine in a binary NS merger of a binary NS. 

\end{abstract}




\section{Introduction}
Since the discovery of gamma-ray bursts (GRBs) in 1963, several lines of observational evidence suggest that the progenitors of short-duration GRBs \citep{2015PhR...561....1K} or nearby short-duration GRBs with extended emission \citep{2022Natur.612..223R,2022Natur.612..232Y,2022Natur.612..228T,2023ApJ...943..146C,2023NatAs...7...67G,2022ApJ...931L..23L,2023A&A...678A.142F, 2024MNRAS.529L..67D, 2024ApJ...962L..27D} originate in the mergers of two compact stellar objects. The leading model is neutron star–neutron star (NS–NS) mergers \citep{1986ApJ...302....1P, 1989Natur.340..126E} or neutron star–black hole (NS–BH) mergers \citep{1991AcA....41..257P}. The first direct detection of the NS–NS merger event GW170817, which was associated with short GRB 170817A and kilonova AT2017gfo, has firmly verified the NS–NS merger model at least for some short GRBs \citep{2017ApJ...848L..12A, 2017ApJ...848L..14G, 2017ApJ...848L..15S, 2018NatCo...9..447Z, 2019MNRAS.486.4479L}. 

Many groups have studied the mergers of two neutron stars (NSs) in numerical relativity simulations \citep{2003MNRAS.345.1077R, 2011ApJ...732L...6R, 2013PhRvD..88d4026H, 2014ApJ...784L..28N}, but the origin of the remnants of NS–NS mergers remains an open question \citep{2011CRPhy..12..206Z}. Since both the NS equation of state and the nascent NS mass are poorly constrained, there are four types of merger remnants that may be formed under certain conditions \citep{2000A&A...360..171R, 2006Sci...311.1127D, 2006MNRAS.372L..19F, 2010CQGra..27k4105R, 2013ApJ...771L..26G, 2013ApJ...763L..22Z, 2014PhRvD..89d7302L, 2014MNRAS.439..744R, 2016PhRvD..93d4065G, 2016PhRvD..94h3010L, 2018MNRAS.480.4402L, 2021GReGr..53...59S}: (1) a black hole (BH) surrounded by a dense torus; (2) a differential-rotation-supported hypermassive NS (HMNS), which may survive for $\sim$300 ms before collapsing into a BH \citep{2003MNRAS.345.1077R, 2012PhRvD..86f4032P}; (3) a supramassive NS (SMNS) supported by rigid rotation, which may survive for hundreds of seconds before collapsing to a BH \citep{2010MNRAS.409..531R, 2017ApJ...835..181L, 2020ApJ...898L...6L}; and (4) a stable NS with a much longer lifetime \citep{2006Sci...311.1127D, 2013ApJ...763L..22Z, 2015ApJ...805...89L, 2016PhRvD..93d4065G}. The dynamical nature of the NS–NS merger suggests that quasi-periodic signals and quasi-periodic oscillations (QPOs) are more probable if the central engine is an HMNS \citep{2019ApJ...884L..16C}, or if internal plateau \footnote{A fairly constant emission followed by a steep decay \citep{2007ApJ...665..599T}.} emission (or plateau emission) in X-rays could be shown at a later time within the SMNS (or stable NS) as the central engine \citep{1998PhRvL..81.4301D, 2001ApJ...552L..35Z}.

From an observational point of view, a small fraction of short GRBs that have plateau (or internal plateau) emission in X-rays point towards a stable NS (or a SMNS) as the central engine of short GRBs \citep{1998PhRvL..81.4301D, 2001ApJ...552L..35Z, 2013MNRAS.430.1061R, 2010MNRAS.409..531R, 2015ApJ...805...89L, 2017ApJ...835..181L}. However, due to their short lifetimes, it is difficult to observe definite clues for the HMNS as the central engine of short GRBs. Recently, \cite{2023Natur.613..253C} discovered kilohertz QPOs in two short GRBs with a confidence level of about $5 \sigma$ from archival Burst And Transient Source Experiment (BATSE) data, and suggest that these are evidence for a shortly surviving HMNS before the final collapse into a BH. 

Motivated by the above-described search in the BATSE data for a QPO that is the signature of an HMNS, this paper presents a systematic analysis of the short-duration GRBs observed with {\em Fermi}/Gamma-ray Burst Monitor (GBM) since its inception in 2008 and investigates whether such QPOs are also hidden in the light curves of those short GRBs. We systematically searched more than 3674 GRBs observed by {\em Fermi}/GBM and found 605 short-duration GRBs. We focused on analysing the 605 short-duration GRBs and found that three short-duration GRBs (e.g. GRBs 120323A, 181222B, and 190606A) with a high-frequency quasi-periodic signal can be identified with a confidence level above 5.2 $\sigma$. Our data reduction and QPO search method are shown in Section 2. The observations and results of our QPO search for those three GRBs are presented in Section 3. Conclusions are drawn in Section 4 with some additional discussions.

\section{Data reduction and QPO search}\label{2}
\subsection{Fermi/GBM data reduction}\label{2.1}

The Fermi satellite was launched in 2008 June and has been in operation for 16 yr. There are two instruments onboard the Fermi satellite. One is the GBM \citep{2009ApJ...702..791M}, which has 12 sodium iodide (NaI) and two bismuth germanate scintillation detectors covering the 8 keV–40 MeV energy band. The other is the Large Area Telescope \citep{2009ApJ...697.1071A}, which has an energy coverage from 20 MeV to 300 GeV. 

In order to test how many short GRBs contain high-frequency quasi-periodic signals, as of 2024 June, we download the GBM data for all GRBs from the public science support centre at the official Fermi website\footnote{$\rm ftp://legacy.gsfc.nasa.gov/fermi/data/gbm/bursts.$}. A PYTHON code based on Fermi GBM DATA TOOLS (v1.1.1)\footnote{$\rm https://fermi.gsfc.nasa.gov/ssc/data/analysis/gbm/$} was developed to extract the light curves for different detectors. We find that the light curves of more than 3674 GRBs can be identified for extraction. More details of light curve extraction can be found in \cite{2011ApJ...730..141Z}. 

We find that the $T_{\rm 90}$ of 605 GRBs out of 3674 GRBs are less than 2 s at the official Fermi website. Here, we focus on studying the 605 short-duration GRBs and extract the light curves of those short-duration GRBs from 12 NaI detectors by adopting a 128 $\mu$s time bin. 

\subsection{QPO search method in short GRB samples}\label{2.2}

The data that we adopted consist of a flux time series of 605 short GRBs, and it includes 12 simultaneous light curves from the 12 NaI detectors in the GBM TTE (time-tagged event) data for each short GRB. We extracted the NaI detector light curve with a time bin of 128 $\mu$s in the energy range of 50–900 keV. Each NaI detector points to a slightly different position on the sky, the light curves of each GRB are ranked in order of increasing angular distance between the GRB position and detector pointing. They are designated as $LC_i$ where $i=1,2...,12$ from closest to farthest. 12 summed light curves are constructed by successively adding one more light curve to the sum in order of the ranking. These are designated as 

\begin{eqnarray}
LC_{\rm sum}(n)=\sum_{i=1}^{n}LC_i , n=1,2,...,12
\end{eqnarray}

These summed light curves are not statistically independent. Next, we applied to $LC_{\rm sum}(n)$ and cut down the light curves to a duration of 2.048 s, which consists of 16,000 bins. Our selection of the start and end times of the light curves can be expressed as 
\begin{eqnarray}
T_{start}=T_{90-mid}-1.024
\end{eqnarray}
\begin{eqnarray}
T_{end}=T_{90-mid}+1.024
\end{eqnarray}
\begin{eqnarray}
T_{90-mid}=\frac{T_{\rm 90-start}+T_{\rm 90-end}}{2}
\end{eqnarray}

Here, the $T_{\rm 90}$ is the duration in which 90 percent of the burst flux was accumulated. The $T_{\rm 90-start}$ is defined as the time at which 5 percent of the total GRB’s flux has been detected, and the $T_{\rm 90-end}$ is defined as the time at which 95 percent of the GRB’s flux has been detected.

The power spectrum of each summed light curve is obtained by calculating the square modulus of the Discrete Fourier transform (DFT; \citep{1970IEEE...18...4B}). More details can be found in \citep{1989ASIC..262...27V,2002MNRAS.332..231U,2012ApJ...746..131B,1983ApJ...266..160L}. The power spectrum $I_{\rm j}$ can be expressed as

\begin{eqnarray}
I_j={\frac{1}{N}}|a_j|^2 , j=1,...,N/2
\end{eqnarray}

Here, the $a_{\rm j}$ is the result of the DFT, and $N$ is the number of bins in a light curve. Each power spectrum $I_{\rm j}$ is fitted with a broken power law plus constant using the maximum likelihood method \citep{2010MNRAS.402..307V} by assuming that the powers are exponentially ($\chi^2_v$ v with 2 degrees of freedom) distributed. The broken power-law function can be expressed as 

\begin{eqnarray}
S_j=\left\{
\begin{array}{cc}
 \beta*j^{-\alpha_1}+C&j\leq j_{b} \\
 \beta*j_{b}^{\alpha_2-\alpha_1}*j^{-\alpha_2}+C&j> j_{b},
\end{array}
\right.
\end{eqnarray}

where $\alpha_1$ and $\alpha_2$ are the exponential slopes, respectively, $j_{\rm b}$ is the break point between the two slopes, and $C$ is the white noise component of the power spectrum. The likelihood function is equivalent to minimizing the function $D$ which is expressed as

\begin{eqnarray}
D=2 \sum_{j=1}^{N/2} \frac{I_j}{S_j}+\ln{S_j}.
\end{eqnarray}
Each power spectrum is divided by its best fit $S_j$, and the renormalization powers are designated as
\begin{eqnarray}
R_j=2\frac{I_j}{S_j},j=1,2,...,8000
\end{eqnarray}

Given the evolution and power spectral leakage of quasi-periodic signals, the signal power may exceed the background noise level at several consecutive Fourier frequencies. Therefore, each power spectrum is searched for excesses in power summed over $k$ consecutive frequency bins for values of $k$ from 1 to 10. This search is performed using the summed power of $8000-k+1$ maximally overlapping groups of powers so these summed powers are not statistically independent. The summed power spectrum $R_k$ can be expressed as

\begin{eqnarray}
R_k(j)=\sum_{j}^{j+k-1}R_j,j=1,2,...,8000-k
\end{eqnarray}

If there are quasi-periodic signals in the power spectrum $R_{\rm k}(j)$, the power of the quasi-periodic signal is likely to be the largest value. This process can be represented as

\begin{eqnarray}
R_{\rm k-max}=Max(R_{\rm k}(j)),
\end{eqnarray}

where $R_{\rm k-max}$ is the largest value in the power spectrum $R_{\rm k}(j)$. Each GRB contains 12 light curves, and we search each power spectrum for 10 values of $k$ and, therefore, obtain 120 $R_{\rm k-max}$ values for each GRB. In order to evaluate whether or not $R_{\rm k-max}$ is generated by quasi-periodic signals, one can calculate the probability that $R_{\rm k-max}$ is not produced by noise. A simulation is performed generating $10^{10}$ power spectra from $10^{10}$ simulated Poisson noise light curves. For each simulated power spectrum, 10 $R_{\rm k-max}$ ($k$=1,...,10) values are generated by the method described above. To estimate a confidence level associated with a given $R_{\rm k-max}$, the number of times $R_{\rm k-max}$ that this $R_{\rm k-max}$ is exceeded in the $R_{\rm k-max}$ simulated power spectra is counted. The probability $G_{\rm k}$ of a given $R_{\rm k-max}$ value is then calculated as 

\begin{eqnarray}
G_{\rm k}=1-\frac{N_{\rm k-max}}{10^{10}}
\end{eqnarray}

where $G_{\rm k}$ is the probability that $R_{\rm k-max}$ is not generated by random noise. Therefore, for each GRB, we can obtain 120 different $G_{\rm k}$ values. If there is a quasi-periodic signal in a GRB, the maximum $G_{\rm k}$ value is the value that is most likely to be produced by the quasi-periodic signal. This procedure can be represented as 

\begin{eqnarray}
G_{\rm max}=max(G_{\rm k})
\end{eqnarray}

However, we note that this maximum probability $G_{\rm max}$ is not the confidence at which any QPO is detected, because the procedure we followed entails searching for the most significant power excess among many statistically dependent trials. The true detection confidence is estimated below in Section 3. We define the probability into four confidence levels, namely 3 $\sigma$(0.997), 4 $\sigma$ (0.99936), 5 $\sigma$ (0.9999994), and 6 $\sigma$ (0.999999998).

A standard analysis to calculate the frequency and error of a quasi-periodic signal is to adopt a Lorentzian function to fit the power spectrum $I_j$ \citep{2023Natur.613..253C}, with the Lorentzian function ($LF_j$) being expressed as

\begin{eqnarray}
LF_j=S_j+L_j=S_j+\frac{A}{1+\frac{(j-v)^2}{(\Delta v)^2}}.
\end{eqnarray}

Here, $S_j$ is the best fit with a broken power-law function, and it is not a free parameter in the fits. $A$ is the power density of the Lorentzian at its center, $v$ is the central frequency of the Lorentzian, and $\Delta v$ is the full width at half-maximum of the Lorentzian. We adopt the central frequency $v$ as the representative frequency of the quasi-periodic signal. By considering the frequency of the quasi-periodic signal which is distributed within a range, we use the 90 per cent integral area of the $L_j$ function within the Fourier frequency as the representative frequency range of the quasi-periodic signal. The lower limit and upper limit of the frequency correspond to the integral area 5 and 95 per cent, respectively.

\section{\textbf{QPO search results in short GRB samples}}
\begin{figure*}
	\includegraphics[scale = 0.3]{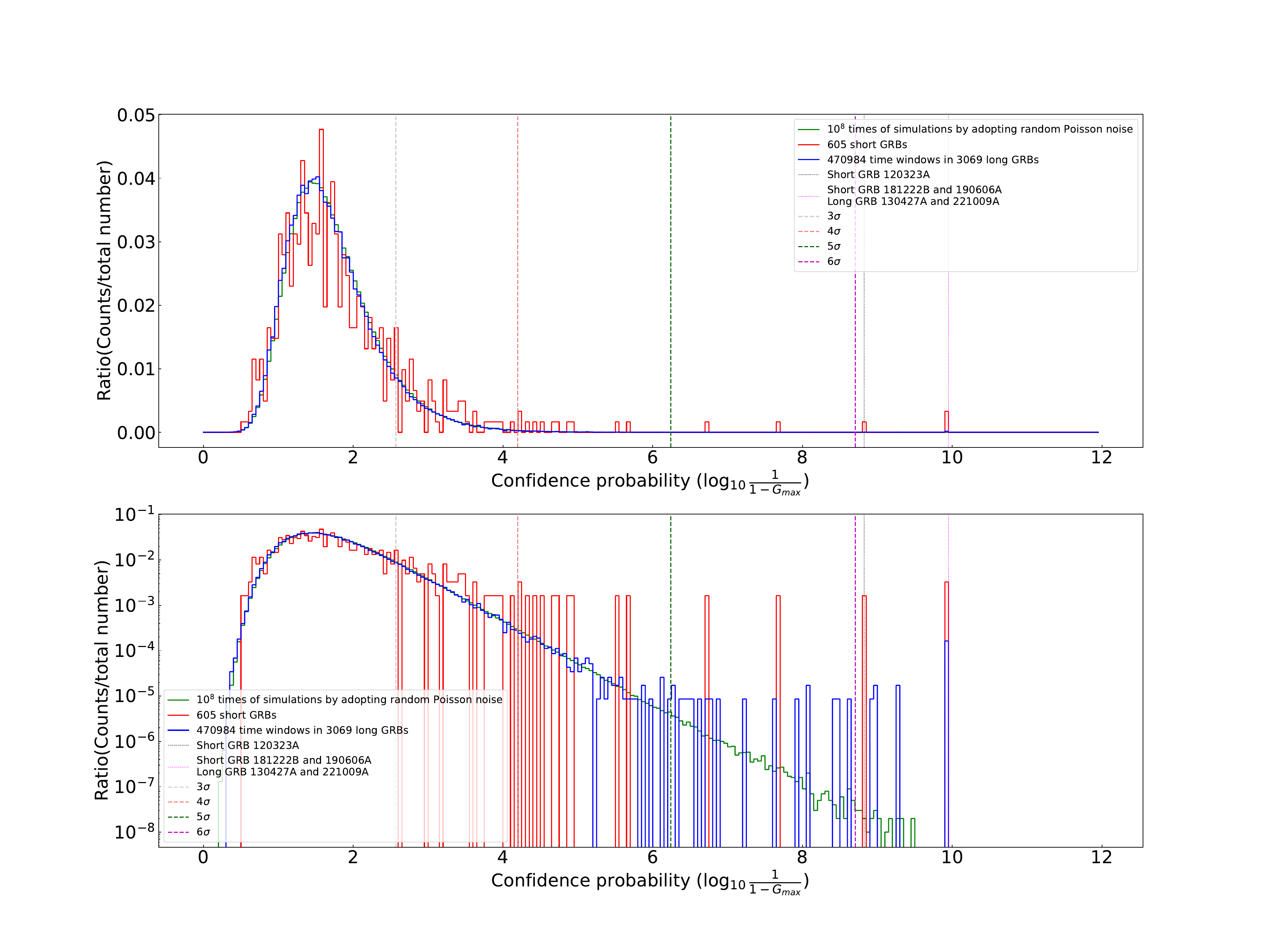}
    \caption{The confidence probability distribution of 605 short GRBs (red line), $10^8$ times of simulations by adopting random Poisson noise (green line), and 470984 time windows in 3069 long GRBs (blue line) with normalization in both linear scale (top) and logarithmic scale (bottom). Different vertical lines correspond to 3 $\sigma$, 4 $\sigma$, 5 $\sigma$, and 6 $\sigma$ of $G_{\rm max}$ values, respectively.}    
    \label{figure:1}
\end{figure*}

Based on the statistical analysis above, one can calculate the power spectrum of each light curve observed by the 12 detectors for a given short GRB. Then we can obtain the $G_{\rm max}$ values for each GRB. Figure \ref{figure:1} shows the $G_{\rm max}$ distribution of 605 short GRBs. We find that there are three short GRBs with $G_{\rm max}$ more than 6 $\sigma$, namely GRBs 120323A, 181222B, and 190606A. The light curves of prompt emission for the three short GRBs are shown in Figure \ref{figure:2}-A. Figure \ref{figure:2}-E shows the power spectra of the three short GRBs, and the horizontal lines are the $R_{\rm k-max}$ values from the $10^{10}$ simulated power spectra (mentioned in Section 2.2) for which $G_{\rm k}$ values is equal to the probability of 3 $\sigma$, 4 $\sigma$, 5 $\sigma$, and 6 $\sigma$. 

We list the fitting results of the Lorentzian function for GRBs 120323A , 181222B and GRB 190606A in Table \ref{tab:1}. The central frequency $v$ is $1258^{+6}_{-6}$ Hz for GRB 120323A, $623^{+4}_{-4}$ Hz for GRB 181222B, and $1410^{+4}_{-5}$ Hz for GRB 190606A, respectively. One question is whether or not such high-frequency QPOs in those three short GRBs are caused by random noise or the instrumental response of the detector. In order to test the possibility of the above effects on the QPO signal, we adopt the simulations to perform test, and the details are described as follows. 

\subsection{The effect of random noise}
Firstly, we adopt random Poisson noise to simulate 12 light curves to represent one GRB with 12 detectors. Each of the simulated 12 light curves are generated and successively added together to produce 12 statistically dependent summed light curves. Then, we calculate the power spectra for each summed light curve. One can obtain 120 $R_{\rm k-max}$ values whose $G_{\rm k}$ values is evaluated by using the results from the earlier $10^{10}$ simulated power spectra. Out of the 120 $G_{\rm k}$ values, the overall $G_{\rm max}$ is selected. Finally, by adopting the same method above, we repeat the simulations $10^8$ times and obtain the distribution of $G_{\rm max}$ of the Poisson noise in this way. The $G_{\rm max}$ distribution of the simulated light curves are shown in Figure \ref{figure:1}. We find that the $G_{\rm max}$ of 13 of the simulations are larger than 6 $\sigma$, it means that there is $1.3\times10^{-7}$ probability to get a $G_{\rm max}$ above 6$\sigma$ in each simulation. Therefore, the confidence level of the three quasi-periodic signals are exceed 5.2 $\sigma$ ( $1.3\times10^{-7}$). Then, we adopt a binomial distribution to calculate the probability of obtaining more than 3 events in our 605 simulations, and the probability is $8.06\times10^{-14}$. If this is the case, the high-frequency QPO what we detected in those short GRBs are not likely to be caused by random noise.

\subsection{The effect of instrumental response}
There are several effects that could result in quasi-periodic signals in the light curve of GRBs such as dead time of detectors, spectral leakage, detection method, and the GRB itself. If the detected quasi-periodic signal in the light curve of a short GRB is not from the burst itself but is caused by the dead time of detectors, then the quasi-periodic signal should exist in other GBM light curves of GRBs as well. In order to test this possibility, we select the 3069 long GRBs $T_{\rm 90}>2$ s observed by Fermi/GBM until June 2024, and search for quasi-periodic signals in the light curves of long GRBs to compare with that of short GRBs. We adopt 3069 long GRBs as the sample to extract light curves of 12 detectors ranging from -20.48 to 307.2 seconds. Because the time window of short GRBs is 2.048 s, we separate each extracted long GRB light curve into 160 time segments with a time window of 2.048 s. If this is the case, the total number of time windows is 470984 for the 3069 long GRBs.

We obtained 21 events with $G_{\rm max}$ values larger than $6 \sigma$ in 470984 time windows. These 21 events all originate in two GRBs, namely GRB 130427A and GRB 221009A. Figure \ref{figure:3} shows the two quasi-periodic signals from GRB 130427A and GRB 221009A. They are extremely bright GRBs that reached the maximum flux limit of the GBM detectors and are reported to be supersaturated in the Fermi GBM \citep{2014Sci...343...42A,2023ApJ...946L..24W}. This means that the prompt emission data of those two GRBs are not intrinsic, so we infer that the detected quasi-periodic signal possibly originated from the detector. The detected fluxes of the three short GRBs (e.g., GRBs 190606A, 120323A, and 181222B) are not supersaturated in the Fermi GBM, which indicates that the detected quasi-periodic signals in the three short GRBs are not likely to be caused by flux supersaturation. The short GRB 181222B maximum flux rate of the second pulse may be nearing supersaturation. But the quasi-periodic signal of GRB 181222B during the first pulse and the rising segment of second pulse (see Section 3.4). Therefore, the quasi-periodic signal in GRB 181222B is not likely caused by flux supersaturation. Moreover, we do not detect other quasi-periodic signals which $G_{\rm max}$ values exceed $6 \sigma$ in the 470984 time windows. We believe that the probability of other  effects causing quasi-periodic signals in the three short GRBs is relatively low. It suggests that the quasi-periodic signal detected in 605 time series is not caused by the dead time of detectors, but is likely to be a unique phenomenon of short GRBs.

On the other hand, we note that there are 13 short GRBs out of 605 with signals which are existence with $G_{\rm max}$ between 4 $\sigma$ and 6 $\sigma$ (see Figure 1). The reason of such signals with high-confidence level that is caused by random noise can be ruled out. It indicates that several short GRBs with quasi-periodic signals out of 13 remain possible to be from the short GRB itself. In order to find out which short GRB with quasi-periodic signal out of 13 is from the short GRB itself, one needs to find out the quasi-periodic signal in those long GRBs with values of $G_{\rm max}$ between 4 $\sigma$ and 6 $\sigma$. However, we cannot determine the origin of quasi-periodic signals in those long GRBs with a value of $G_{\rm max}$ between 4 $\sigma$ and 6 $\sigma$. Therefore, we cannot confirm whether the quasi-periodic signals in short GRBs are different from those in long GRBs. So that, we only discuss those short GRBs with $G_{\rm max}$ values exceeding 6 $\sigma$ in this paper.

\subsection{The effect of detected method}
\begin{figure*}
\centering
	\includegraphics[scale = 0.15]{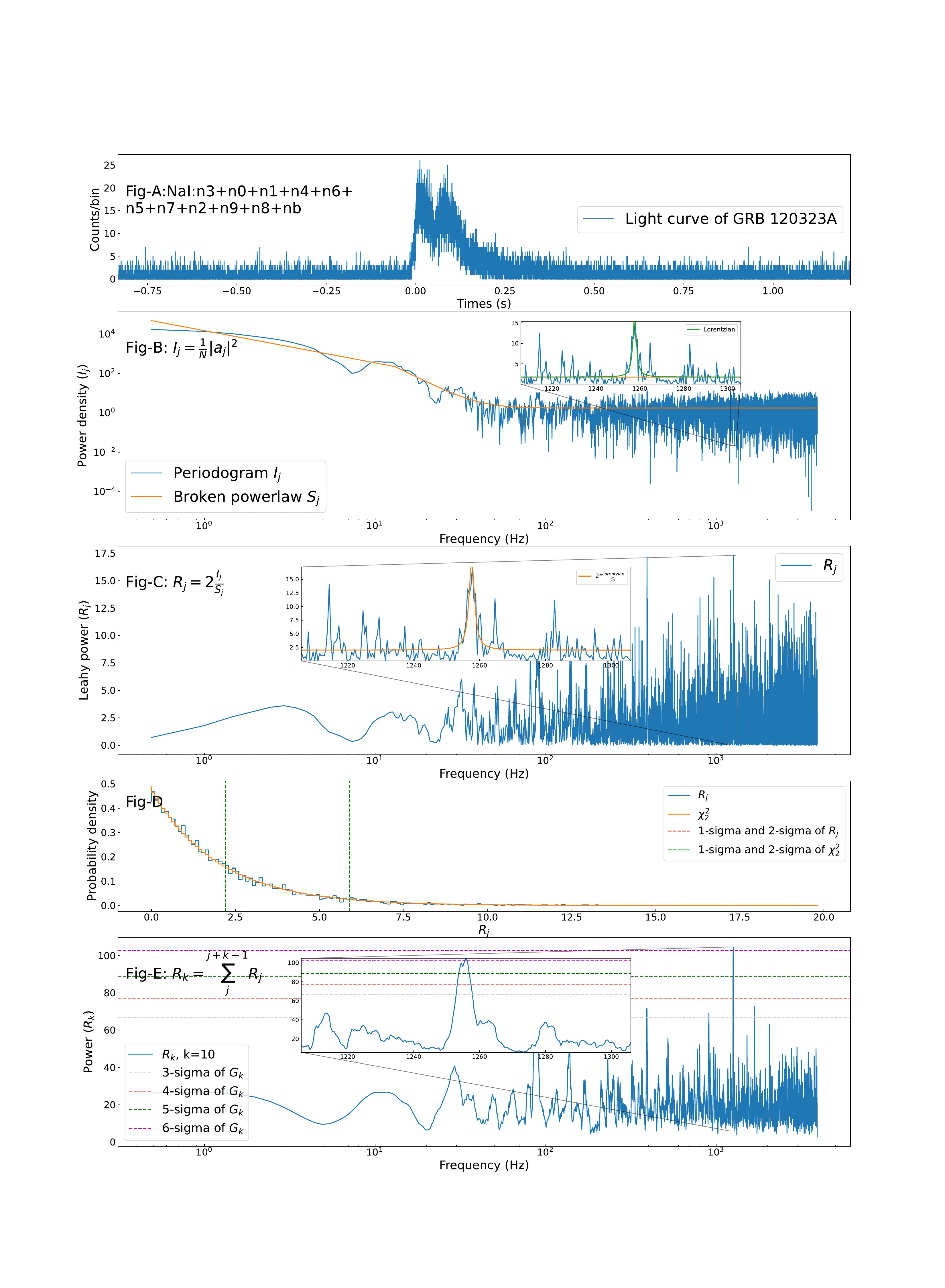}
        \includegraphics[scale = 0.15]{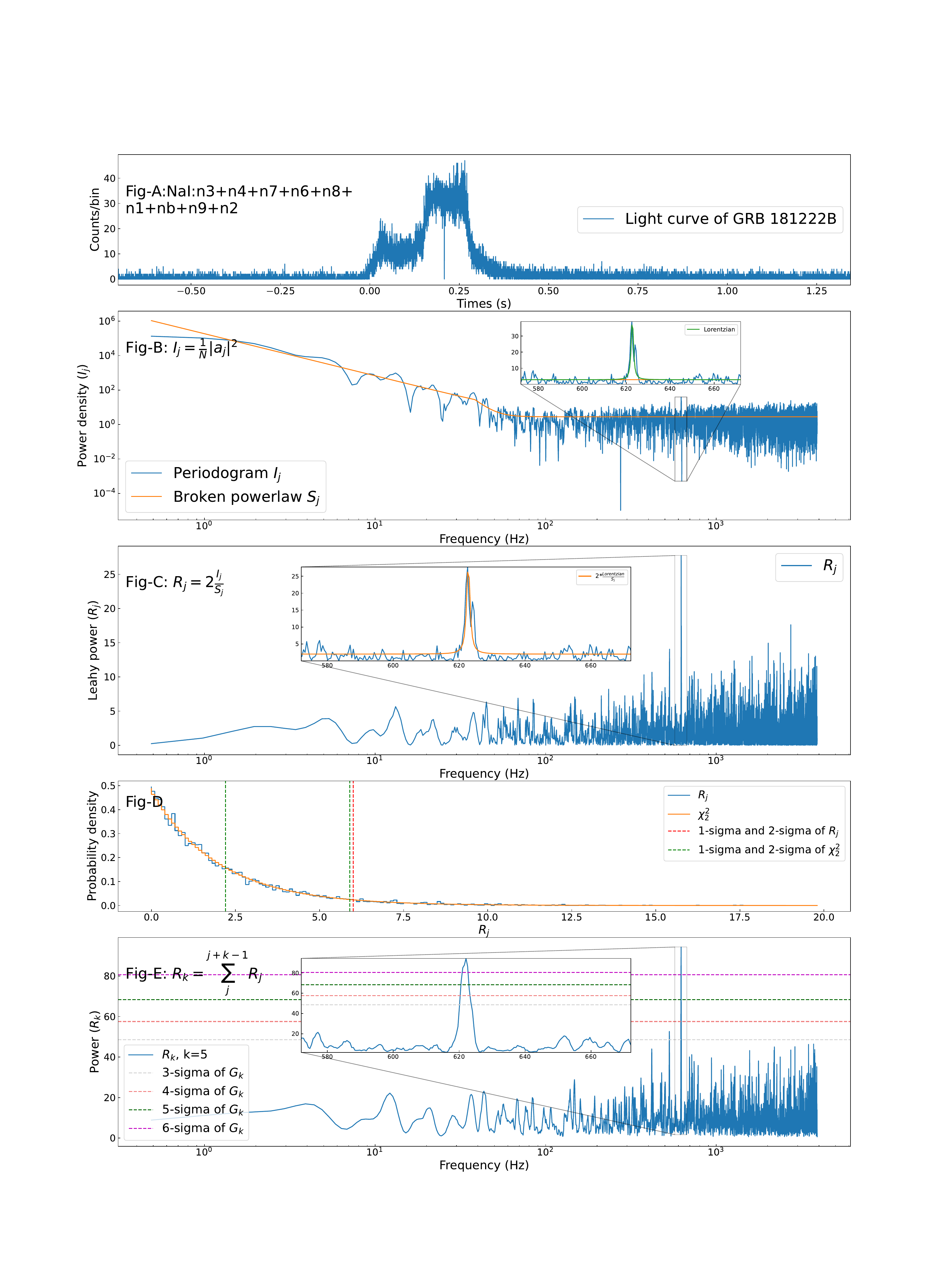}
        \includegraphics[scale = 0.15]{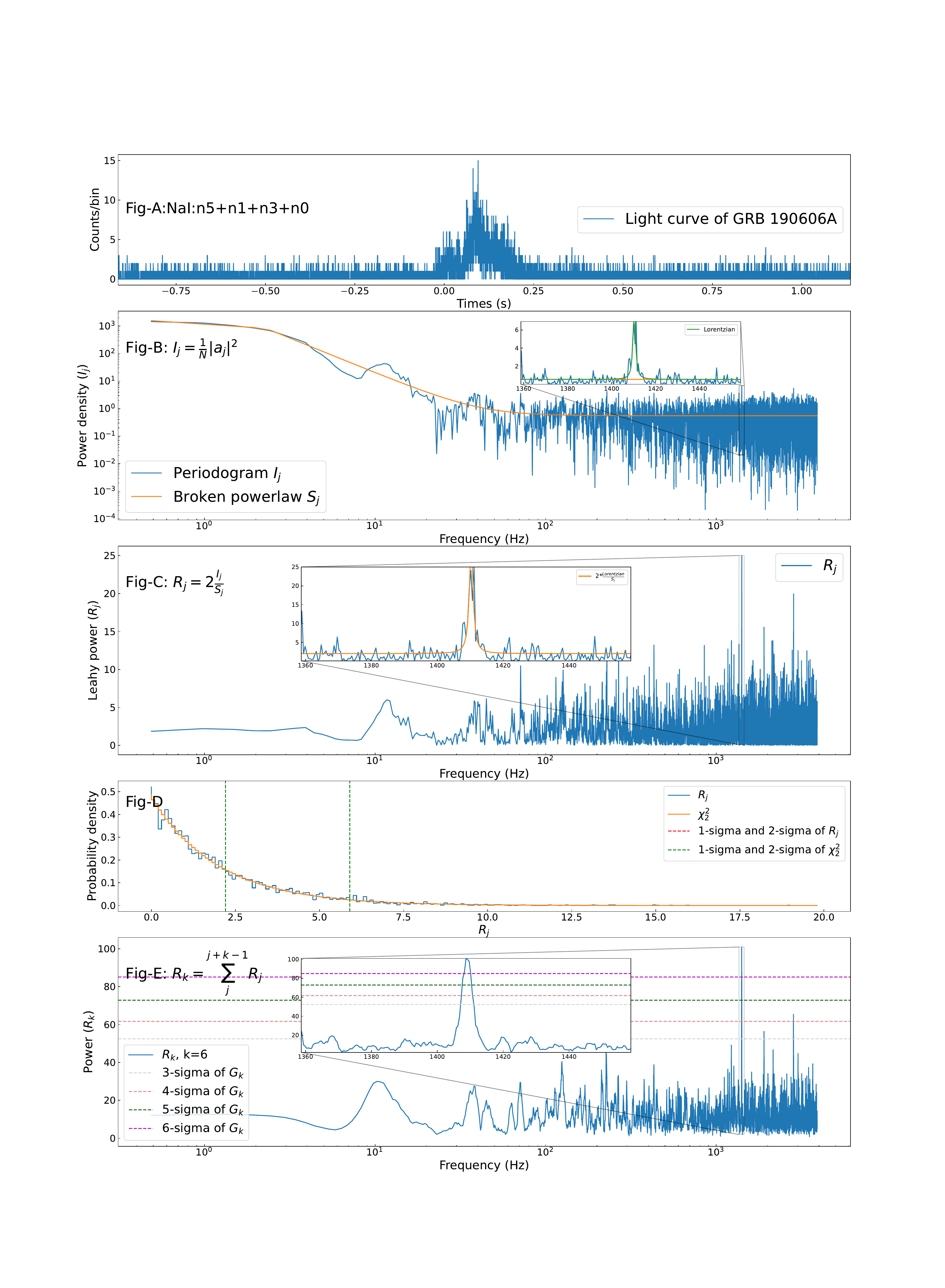}
    \caption{The merged light curves of prompt emission of short GRBs 120323A, 181222B, and 190606A (A). Power spectrum $I_{\rm j}$ and $S_{\rm j}$ with broken power-law fit(B) and Lorentzian function in inset window. The ratio $R_{\rm j}$ between $I_{\rm j}$ to $S_{\rm j}$(C). Distributions of $R_{\rm j}$ and $\chi^2_2$ (D), and $R_k(j)$ (E). Confidence levels shown are for the value of k indicated in each figure, as determined from simulations (see Section 3).}
    \label{figure:2}
\end{figure*}

\begin{figure*}
\centering
	\includegraphics[width=\columnwidth]{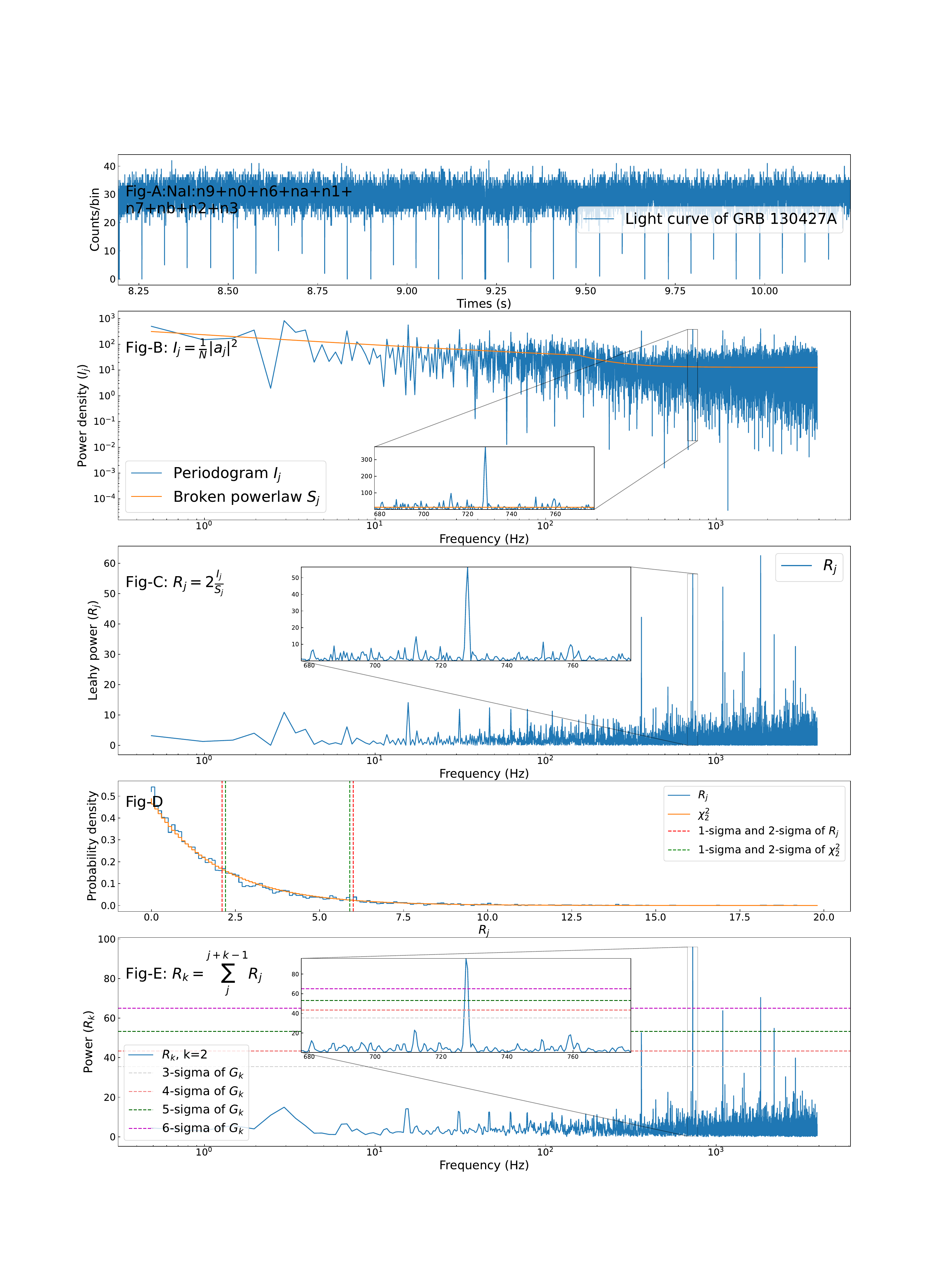}
        \includegraphics[width=\columnwidth]{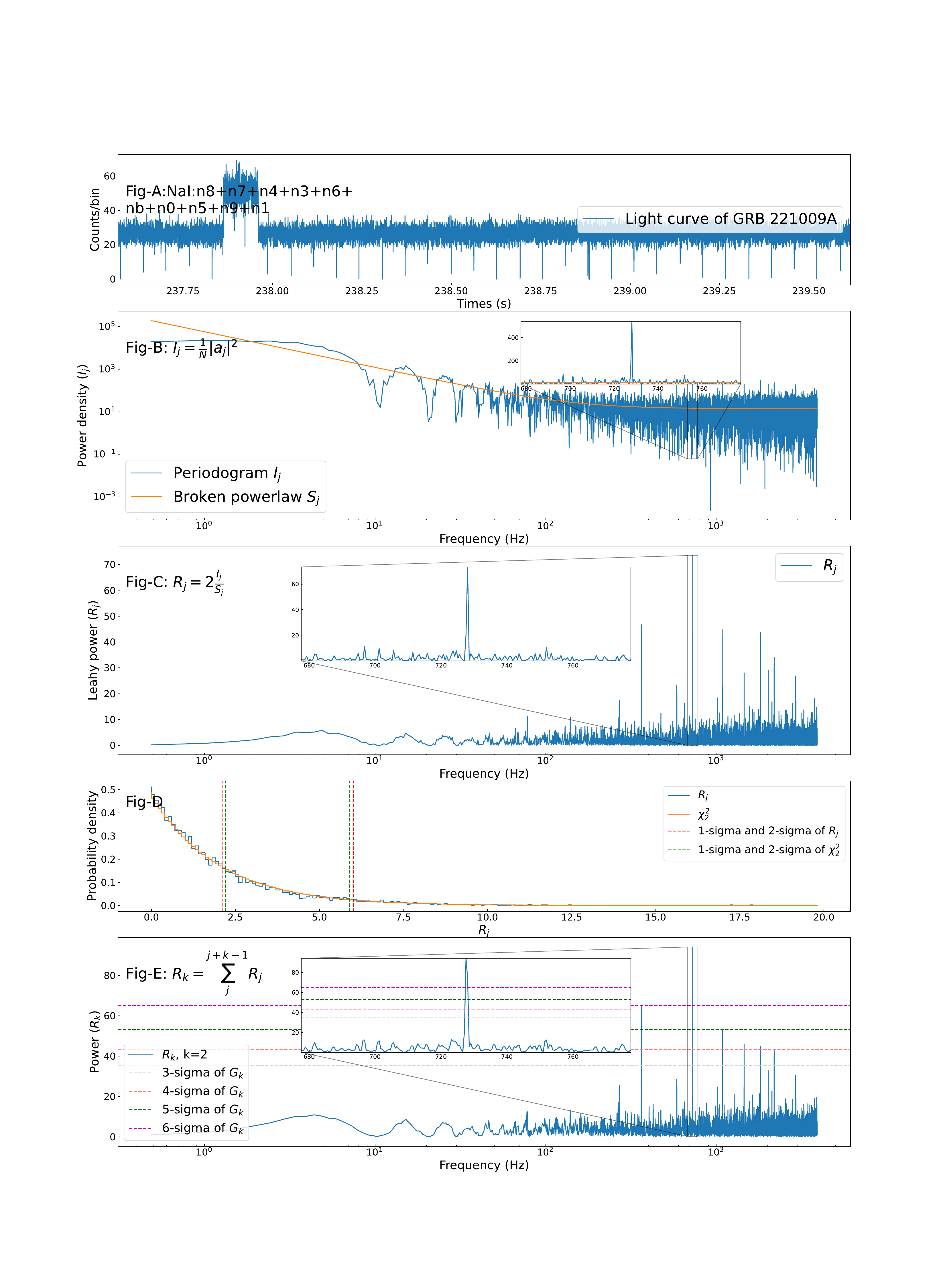}
    \caption{The same as Figure \ref{figure:2}, but for GRB 130427A (left) and GRB 221009A (right).} 
    \label{figure:3}
\end{figure*}

Excepting the effects from the random noise and instrumental response, the detection method of a quasi-periodic signal can also affects the confidence level of the quasi-periodic signal (such as spectral leakage). The ratio $R_{\rm j}$ will only follow the chi-squared distribution with 2 dof expected from a random process time series if the denominator $S_{\rm j}$ correctly describes the mean of the parent distribution from which the $I_{\rm j}$ was drawn (see Figure \ref{figure:2}-B). This is due to the fact that small deviations in $S_{\rm j}$ from the correct value strongly affect the statistics of any power. One needs to check whether the photon noise in the light curve is Poisson noise or not, and provide the best $S_{\rm j}$ fitting results.

\begin{figure*}
\centering
	\includegraphics[width=\columnwidth]{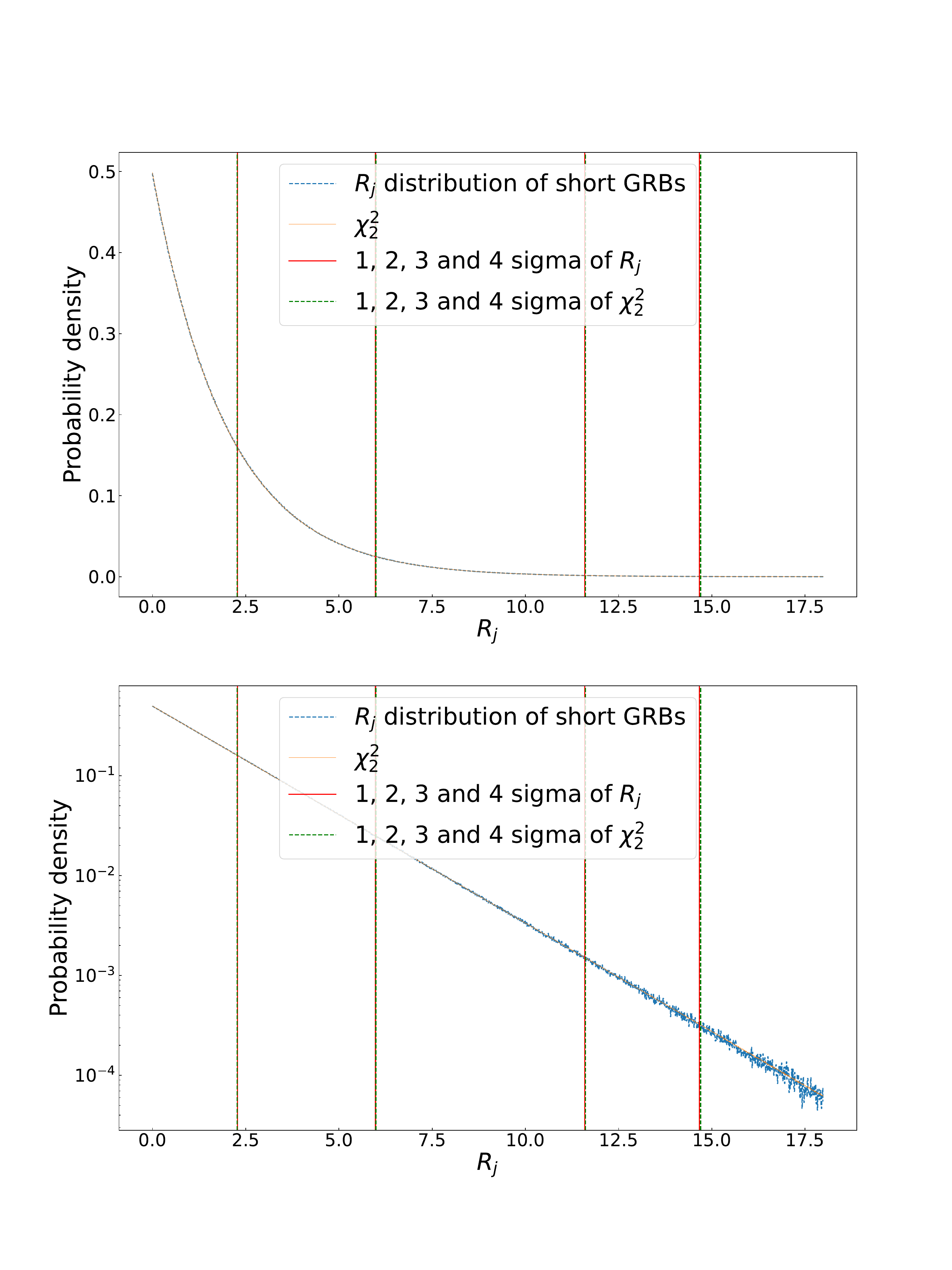}
        \includegraphics[width=\columnwidth]{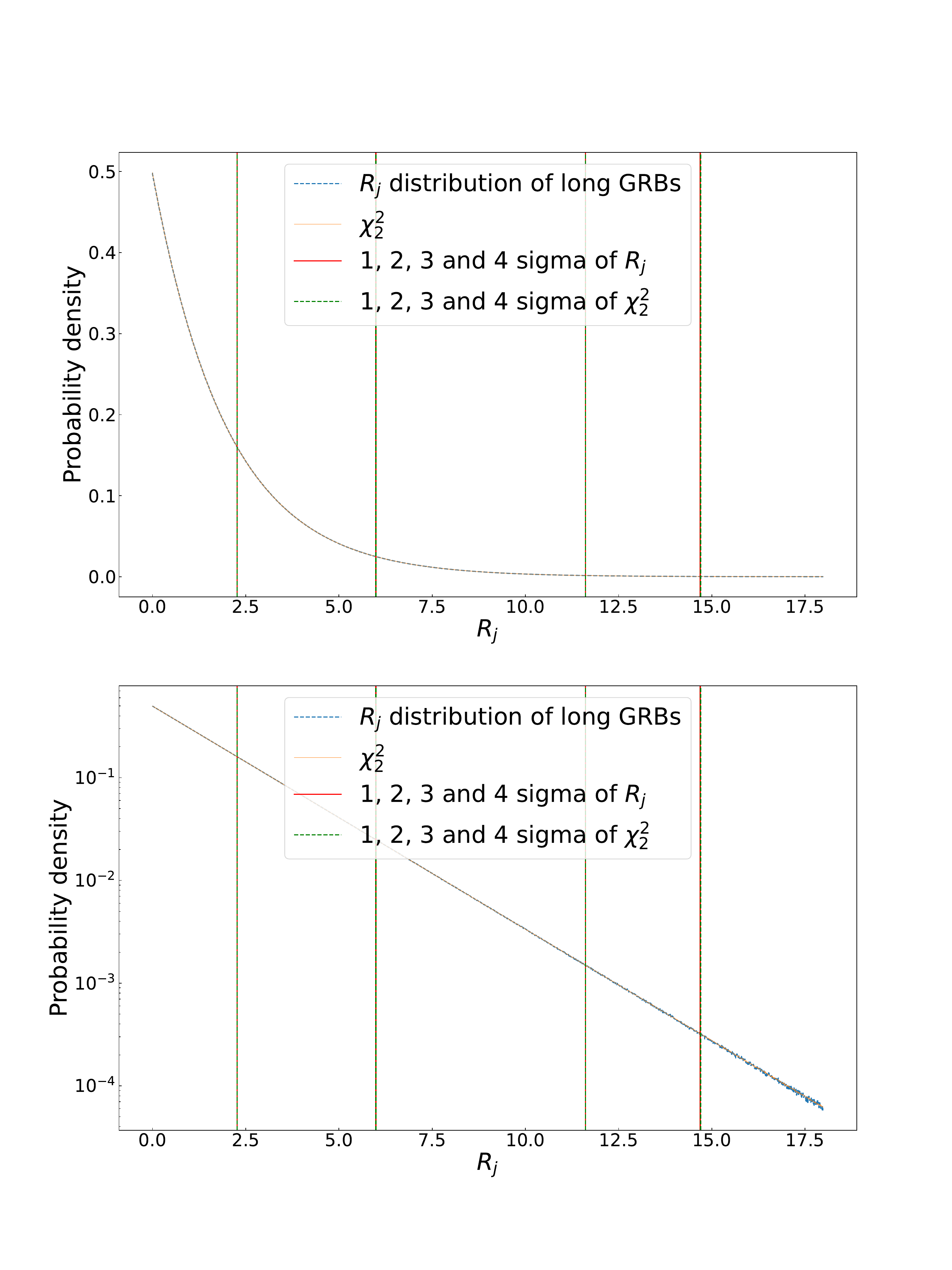}
    \caption{Distributions of all $R_j$ values of short GRBs (left) and Long GRBs (right) in both linear scale (top) and logarithmic scale (bottom), and $\chi^2_2$ distribution.}
    \label{figure:4}
\end{figure*}

\begin{figure*}
\centering
	\includegraphics[scale = 0.2]{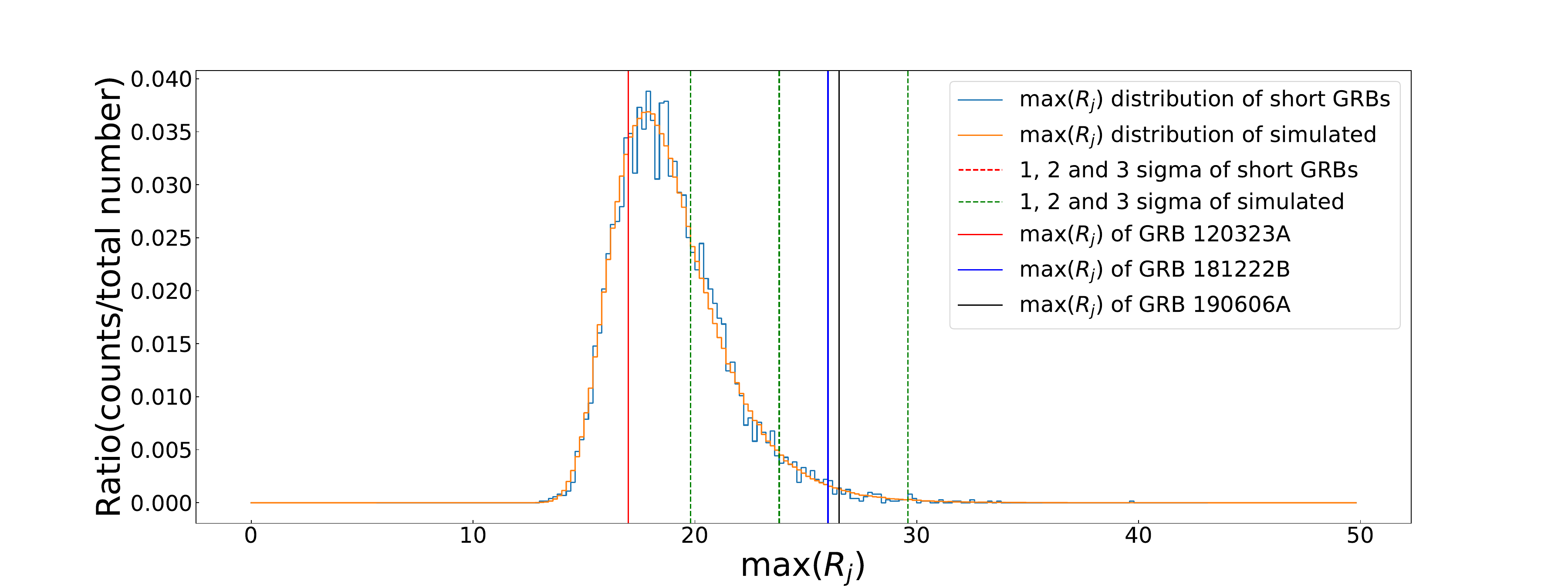}
        \includegraphics[scale = 0.2]{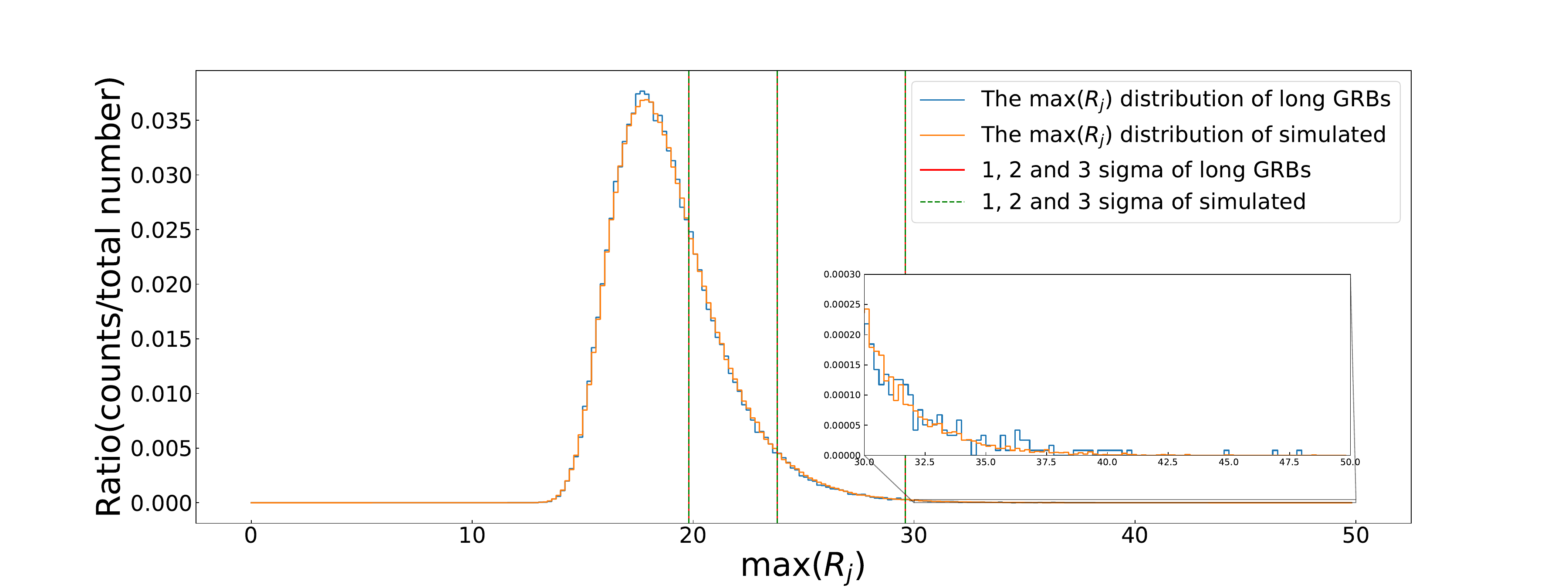}
    \caption{Distributions of $\rm max(R_j)$ values of short GRBs (top), and Long GRBs (bottom), and $10^8$ simulated GRBs.}
    \label{figure:5}
\end{figure*}

In Figure \ref{figure:4}, we show the $R_{\rm j}$ distributions of all power spectra for short GRBs and long GRBs, respectively. The distributions of $R_{\rm j}$ for both short GRBs and long GRBs are almost consistent with the chi-square distribution. If the denominator $S_{\rm j}$ does not correctly describe the mean of the parent distribution from which the $I_{\rm j}$ were drawn, the maximum $R_{\rm j}$ values of both short GRBs and long GRBs in the power spectrum should be overestimated. We show the distributions of the maximum $R_{\rm j}$ values of both short GRBs and long GRBs, and then compare them with the simulated $10^8$ GRBs in Figure \ref{figure:5}. The maximum $R_{\rm j}$ of short GRBs 120323A, 181222B and 190606A are also shown in Figure \ref{figure:5}. It is found that the distributions of the maximum $R_{\rm j}$ values of both short GRBs and long GRBs are similar to that of simulated results, and the confidence level of short GRBs and long GRBs within 3 $\sigma$ completely overlap with that of the simulated GRBs. This indicates that the majority of the fitting results of short GRBs and long GRBs follow a chi-square distribution with 2 dof.

\begin{figure*}
\centering
	\includegraphics[width=\columnwidth]{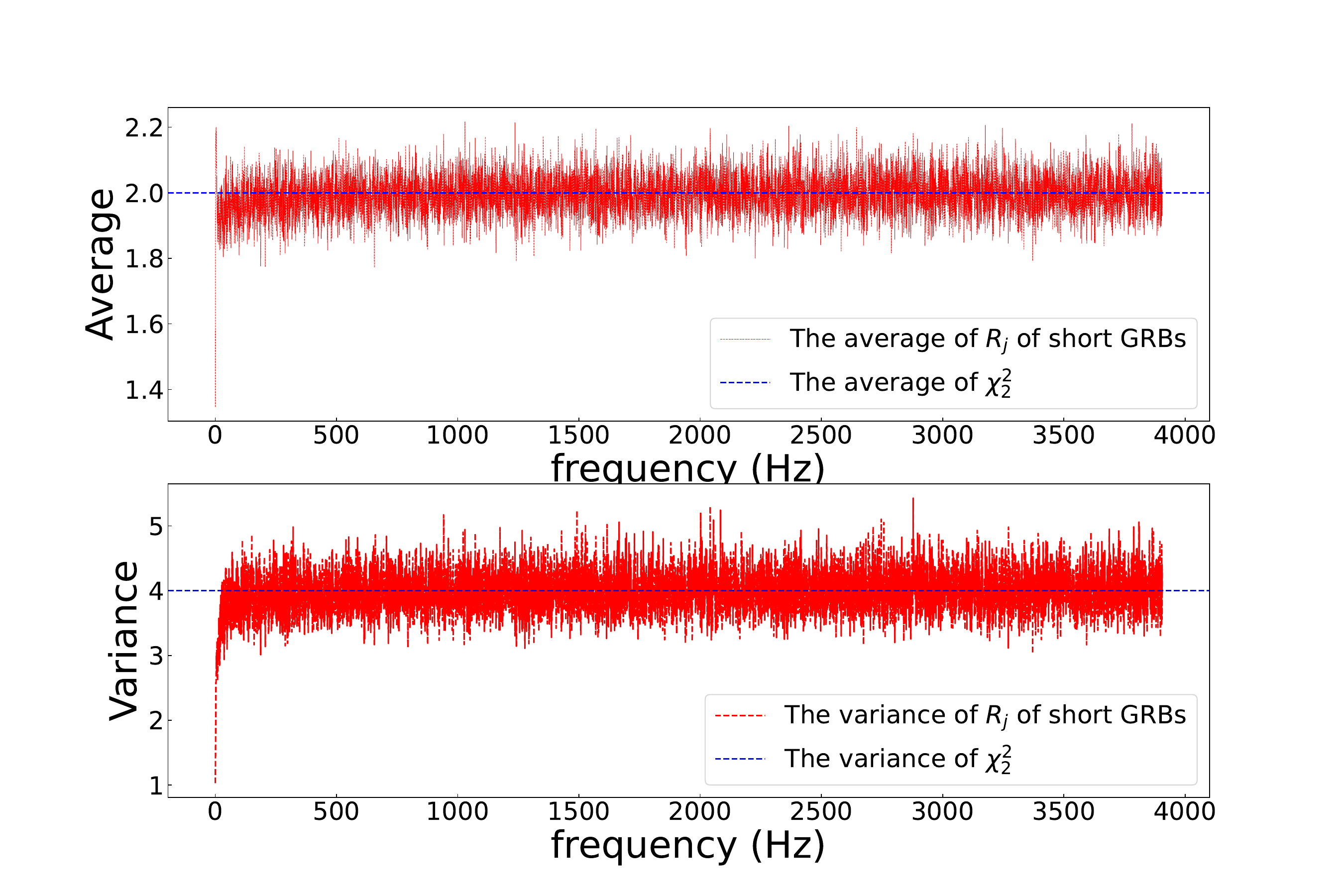}
        \includegraphics[width=\columnwidth]{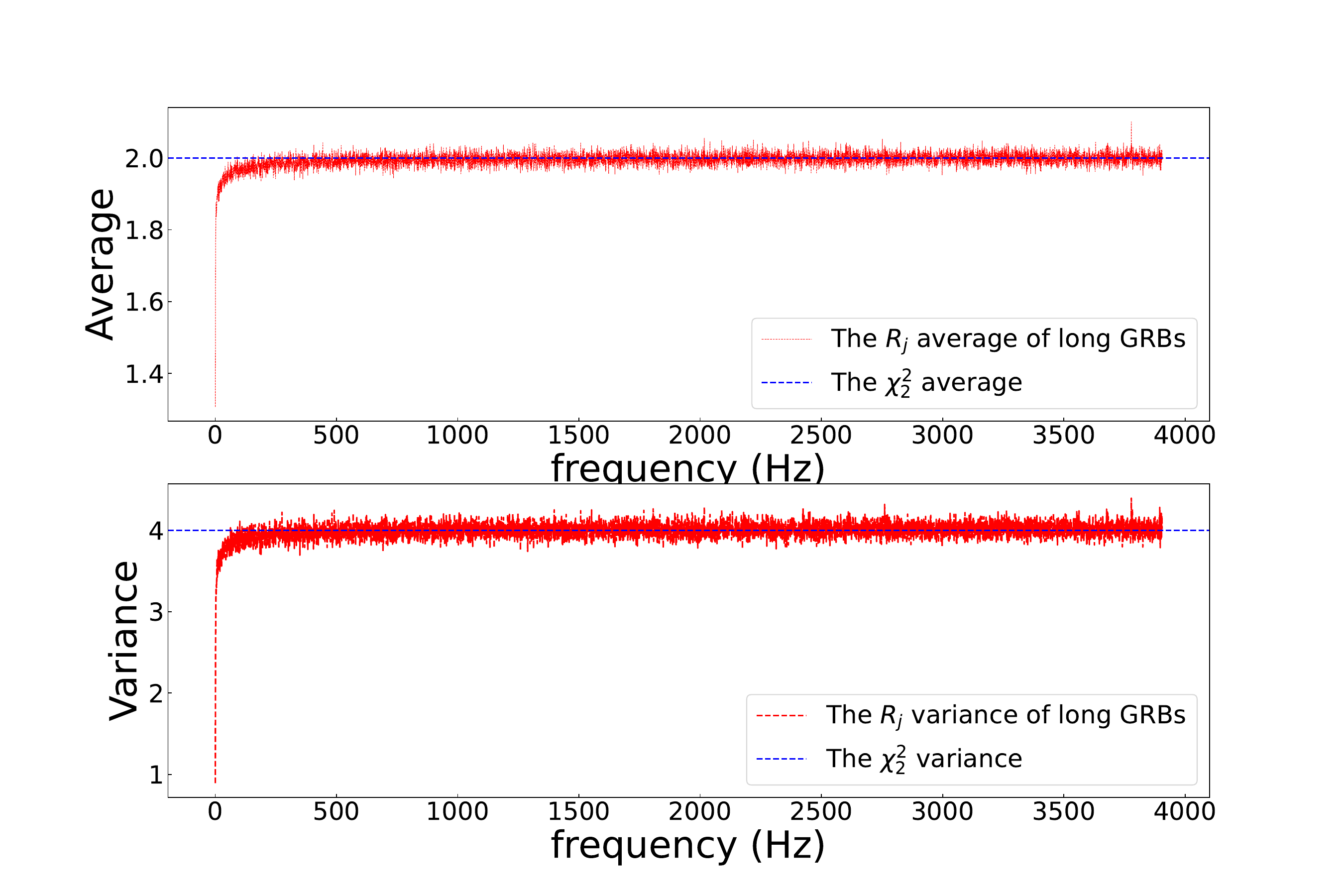}
    \caption{The average and variance of $R_{\rm j}$ values of each Fourier frequency for short GRBs (left) and Long GRBs (right). The blue dashed lines are average and variance of $\chi^2_2$ distribution.}
    \label{figure:6}
\end{figure*}

On the other hand, in order to check whether the $R_{\rm j}$ distributions of short GRBs and long GRBs are dependent on frequency or not, we show the average and variance of $R_{\rm j}$ as a function of frequency in Figure \ref{figure:6}. The average and variance of the chi-square distribution with 2 dof are 2 and 4, respectively. We find that the average and variance values of $R_{\rm j}$ at frequencies above 200 Hz do not significantly deviate from the results of the chi-square distribution with 2 dof (see Figure \ref{figure:6}). However, the average and variance values of $R_{\rm j}$ are underestimated at frequencies below 200 Hz by comparing with the chi-square distribution with 2 dof (see Figure \ref{figure:6}). The underestimated $R_{\rm j}$ values may result in a lower confidence level, but it does not affect a signal with a higher confidence level.

Finally, based on Figure \ref{figure:2}-B and Eq.(8), one can plot the power spectrum of $R_{\rm j}$ in Figure \ref{figure:2}-C. Moreover, one needs to examine whether $R_{\rm j}$ is overestimated in GRBs 190606A, 120323A, and 181222B, so we add the distribution of $R_{\rm j}$ in Figure \ref{figure:2}-D. It is found that the $R_{\rm j}$ distributions in GRB 190606A, 120323A, and 181222B are similar to that of the chi-square distribution with 2 dof, and the 1 $\sigma$ and 2 $\sigma$ lines are coincident with that of the chi-square distribution. Therefore, the quasi-periodic signals in the three short GRBs are not overestimated.

\subsection{The duration of the quasi-periodic signal}
\begin{figure*}
\centering
	\includegraphics[width=\columnwidth]{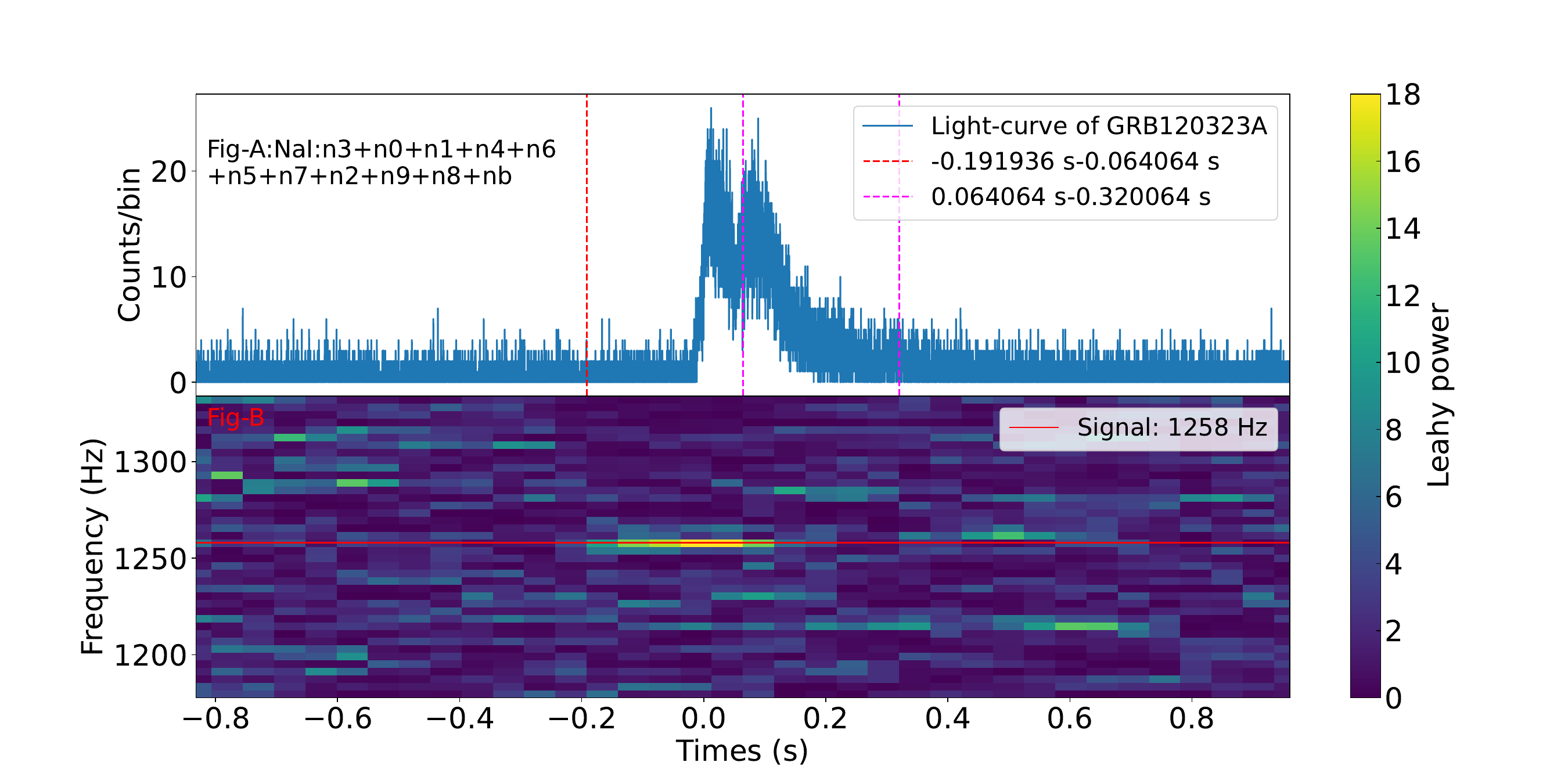}
        \includegraphics[width=\columnwidth]{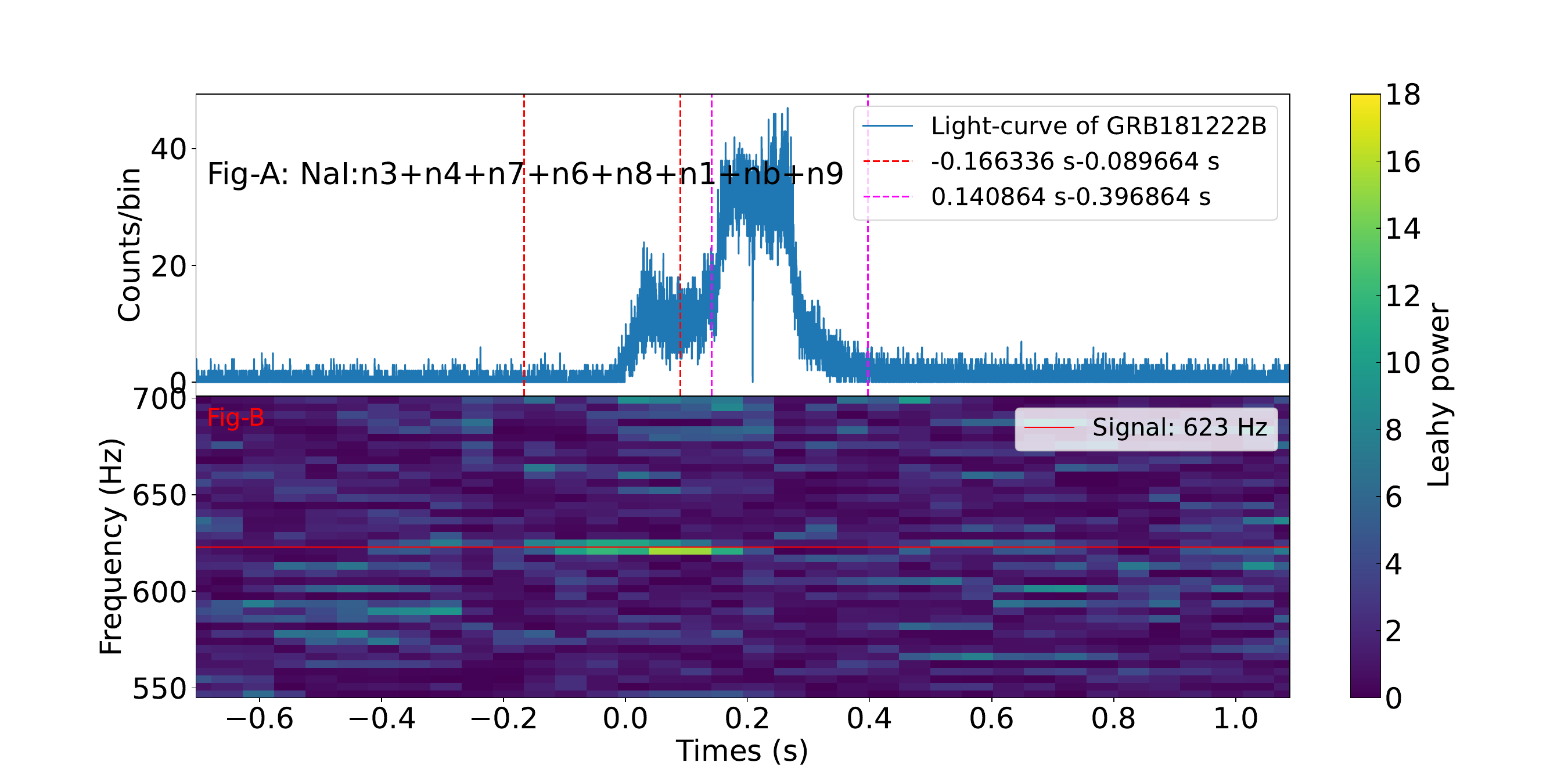}
        \includegraphics[width=\columnwidth]{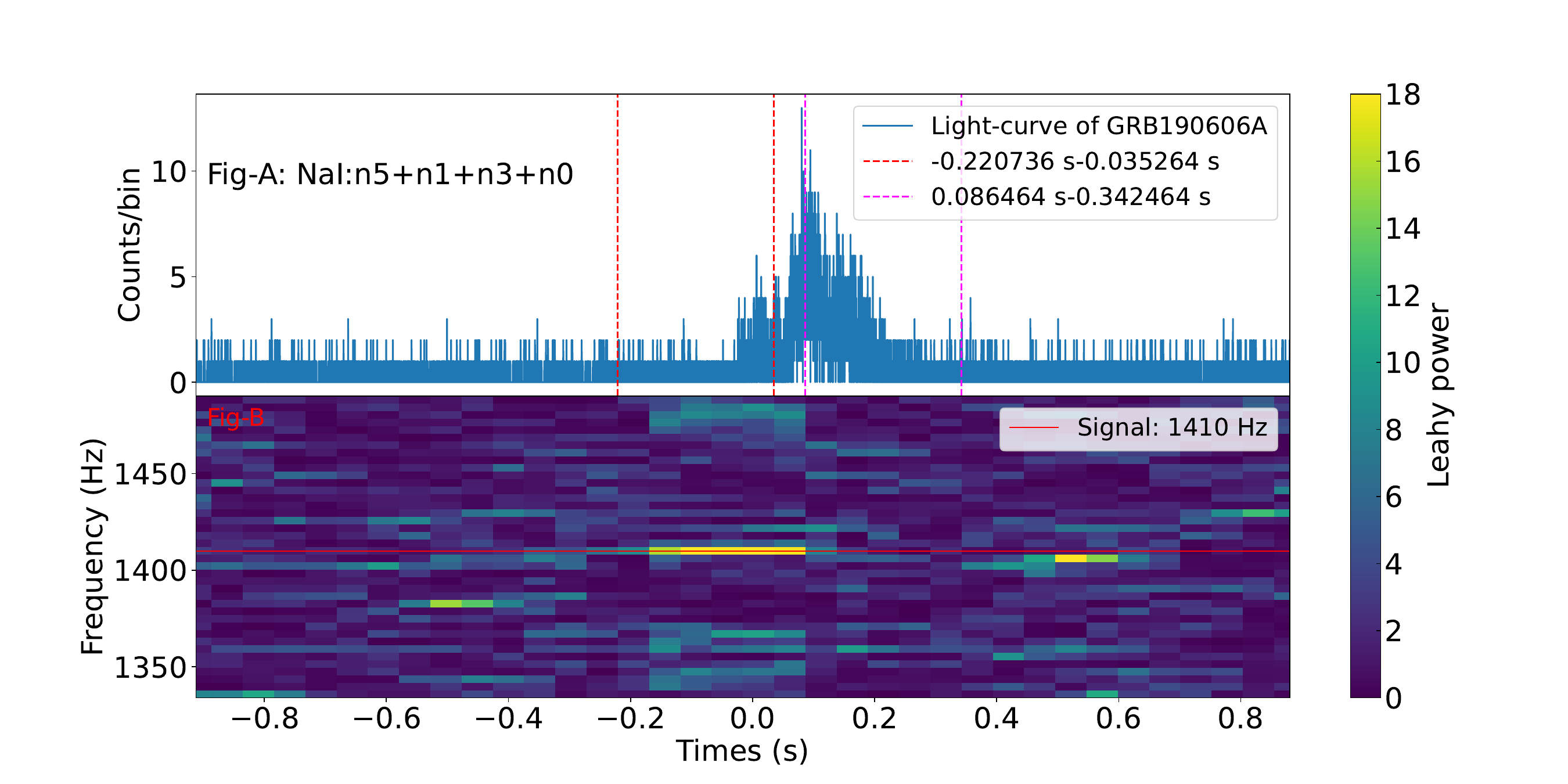}
    \caption{Light curves of prompt emission of short GRBs 120323A, 181222B, and 190606A (A). 2D image-style graph of power spectrum in 0.256 time window with step size of 0.0512 seconds (B). The red dashed and purple dashed lines are the first and last time segments where the signal exceeds 2-sigma, respectively. The powers corresponding to each 0.256 s time window are plotted in the first 0.0512 s of that window.}
    \label{figure:7}
\end{figure*}

Another important question is whether the quasi-periodic signal exists during the prompt emission of those three short GRBs. One needs to understand the duration of the quasi-periodic signal with a high confidence probability. Figure (\ref{figure:7}) shows the light curves of prompt emission (A) and two-dimensional image-style graph of power spectrum (B). The image-style graph of the power spectrum in Figure \ref{figure:7}(B) is composed of a time window of 0.256 seconds with a step size of 0.0512 seconds. It indicates that at least five time windows are shared data for any length of signal, so we only focus on the duration of the signal which is more than five consecutive time windows. One can obtain the minimum ($T_{\rm min}$) and maximum ($T_{\rm max}$) duration of the signal when the number of consecutive time windows of the signal ($N_{\rm SEG}$) exceeds five, namely,  
\begin{eqnarray}
T_{\rm min}=(N_{\rm SEG}-6)*0.0512
\end{eqnarray}
\begin{eqnarray}
T_{\rm max}=(N_{\rm SEG}-5)*0.0512
\end{eqnarray}
For example, the first time windows do not overlap exactly with that of the last time series if the number of consecutive time windows in which the signal is detected is equal to 6, namely, 6 time steps of 0.0512 s.

Next, we consider the power spectrum $R_{\rm j}$ to be greater than 6 as the standard for the existence of a signal. The probability of exceeding a value of 6 for a chi-squared distribution with 2 dof is 0.05, which is approximately 2 $\sigma$. Based on the results of the analysis, we find that there are 7 time windows in both GRB 181222B and GRB 190606A that detected the signal, and 6 time windows in GRB 120323A with a signal detection. The first and last time windows of those three short GRBs are included in its prompt emission (see Figure \ref{figure:7}(A)). Therefore, we believe that the detected high-frequency quasi-periodic signals are likely to originate in the prompt emission of the short GRBs themselves.

\section{Conclusions}
The remnants of NS-NS mergers remain an open question, and the solution depends on the poorly constrained equation of state. One possible remnant for such a merger may be a HMNS which is supported by differential rotation that may survive for about 300 ms before collapsing into a black hole. If this is the case, it may leave a clue with a high-frequency QPO in the prompt emission of short GRBs. In this work, by systematically searching for more than 3674 GRBs (including 605 short GRBs) observed by Fermi/GBM, we find three short GRBs (e.g., GRBs 120323A, 181222B, and 190606A) with QPO that peaked at $1258^{+6}_{-6}$ Hz, $623^{+4}_{-4}$ Hz, and $1410^{+4}_{-5}$ Hz, respectively, all with a confidence level above 5.2 $\sigma$.

Based on the data analysis of our QPO search for those three short GRBs, we find that the duration of a detected QPO is shorter than that of the burst. For example, the start and end times of the burst for GRBs 120323A, 181222B, and 190606A are [0, 0.384] s, [0.032, 0.576] s, and [0, 0.224] s, respectively. However, the start and end time of the detected QPO for those three short GRBs are [0.0384$\pm$0.0256, 0.0896$\pm$0.0256] s, [0.064$\pm$0.0256, 0.1664$\pm$0.0256] s, and [0.01$\pm$0.0256, 0.112$\pm$0.0256] s, respectively. On the other hand, the light curves of those three short GRBs seem to be composed of two pulses, and it is found that the QPO signals exist during the first pulse and the rising segment of the second pulse, but disappear during the decay segment of the second pulse. In general, two effects may result in such a situation. One is incorrect data analysis with the selected time-step, but we adopted a time-step of duration 0.0512 s which is much less than the burst duration. This smaller time-step is not likely to result in such a situation. The QPO signal disappearing during the decay segment of the second pulse may arise in the intrinsic physics, but this remains debatable.

From the theoretical point of view, a HMNS with ultra-strong magnetic fields induced by small-scale turbulent processes that formed via neutron star mergers may be a potential source producing the high-frequency QPO in short GRBs \citep{2006Sci...312..719P, 2015PhRvD..92l4034K, 2019ApJ...884L..16C, 2023ApJ...947L..15M}. After the HMNS forms, the strong toroidal magnetic fields in the surface of the star induce buoyant instabilities \citep{1978RSPTA.289..459A} which can then lead to a rise of poloidal loops in the surface layers \citep{2011A&A...532A..30K}. If this is the case, a strong differential rotation is naturally formed to inflate these loops \citep{2017PhRvD..96d3004H, 2016PhRvD..94d4060K}, and then launch quasi-periodic emissions of powerful electromagnetic radiation \citep{2014ApJ...788...36B}. Recently, numerical relativity simulations suggest that the quasi-periodic electromagnetic substructure is dominated by magnetohydrodynamic shearing, which may naturally explain the recently reported QPO in the short GRB 910711 \citep{2023Natur.613..253C, 2023ApJ...947L..15M}. If a HMNS is indeed operating in some GRBs, searching for such high-frequency QPO in the short GRBs is very important for understanding the NS equation of state \citep{2015PhRvD..92l1502P, 2022PhRvL.128p1102B}.

However, invoking an HMNS to interpret the high-frequency QPO in the prompt emission of those three short GRBs (e.g. GRBs 120323A, 181222B, and 190606A) imposes two curious questions that are not clarified in either theory or numerical relativity simulations of NS mergers. One question regards the composition and structure of an HMNS. Initially, the temperature should be high enough shortly after the formation of the HMNS, and the composition of the HMNS should be dominated by fluid instead of rigid structure because there has not been enough time to cool down. If this is the case, the oscillation of fluid seems to have difficulty producing such high-frequency quasi-periodic signals. The other question is how to produce the QPO signals only in the first pulse and the rising segment of the second pulse. There remains a poor understanding of this question. 

In order to investigate the remnants of those three short GRBs-like events and find evidence for an HMNS in a double NS merger, more information from the simultaneous detection of both gravitational wave and multi-band electromagnetic radiation is essential. We therefore strongly encourage follow-up observations when the LIGO Virgo–KAGRA (LVK) Collaboration detects the gravitational wave signal of those three short GRBs-like events in the future.

\section*{Acknowledgements}
We acknowledge the use of the public data from the Fermi/GBM. This work is supported by the Guangxi Science Foundation the National (grant No. 2023GXNSFDA026007), the Natural Science Foundation of China (grant Nos. 12494574, 11922301 and 12133003), the Program of Bagui Scholars Program (LHJ), and the Guangxi Talent Program (“Highland of Innovation Talents”).

\section*{Data Availability}
The data what we adopt is publicly available data from Fermi/GBM, and there are no new data associated with this article.










\bsp	

\clearpage

\begin{table*}
 \caption{The best-fitting parameters of the Lorentzian function}
 \label{tab:1}
 \begin{tabular*}{\textwidth}{@{\hspace*{20pt}}l@{\hspace*{55pt}}l@{\hspace*{55pt}}l@{\hspace*{55pt}}l@{\hspace*{55pt}}l@{}}
  \hline
  GRB & $v$ (Hz) & $\Delta v$ (Hz)& power density of Lorentzian $A$ & frequency range (Hz)\\
  \hline
  120323A & 1258 & 1 & 14.44 & $1252-1264(1258^{+6}_{-6})$\\[3pt] 
  181222B & 623 & 0.6 & 37.04 & $619-627(623^{+4}_{-4})$\\[3pt]
  190606A & 1410 & 0.7 & 7.09 & $1405-1414(1410^{+4}_{-5})$\\[3pt]
  \hline
 \end{tabular*}
\end{table*}

\end{document}